%% file: promotweet.tex
\documentclass{sig-alternate-10pt}

\usepackage{url}
\usepackage{amsmath}
\usepackage{multirow}
\usepackage{graphicx}
\usepackage{epstopdf}
\usepackage{color}

\def\etal{{\it et al.}}
 
\begin{document}
\title{Topic and Sentiment Analysis on OSNs: \\a Case Study of Advertising Strategies on Twitter}

\numberofauthors{1}
\author{
\alignauthor Shana Dacres, Hamed Haddadi, Matthew Purver\\
        \affaddr{Cognitive Science Research Group}\\     
         \affaddr{School of Electronic Engineering and Computer Science}\\           
        \affaddr{Queen Mary University of London, UK}\\  
        \affaddr{firstname.lastname@eecs.qmul.ac.uk}\\  
}

\maketitle

\begin{abstract}
\input{abstract}
\end{abstract}




\input{intro}

\input{related}

\input{data}

\input{classify}
\input{results}

\input{conclusion}

%
\section*{Acknowledgments}
The authors thank Chatterbox\footnote{\url{http://chatterbox.co}} for providing unlimited access to the Sentimental API.

\vspace{0.3cm}
 \bibliographystyle{abbrv}
 \bibliography{adverts}  
 \balancecolumns

\end{document}

%% file: abstract.tex
Social media have substantially altered the way brands and businesses advertise:
Online Social Networks provide brands with more versatile and dynamic channels
for advertisement than traditional media (e.g., TV and radio). Levels of
engagement in such media are usually measured in terms of content adoption
(e.g., \emph{likes} and \emph{retweets}) and sentiment, around a given
topic. However, sentiment analysis and topic identification are both non-trivial
tasks.

In this paper, using data collected from Twitter as a case study, we analyze how
engagement and sentiment in promoted content spread over a 10-day period. We
find that promoted tweets lead to higher positive sentiment than promoted
trends; although promoted trends pay off in response volume. We observe that
levels of engagement for the brand and promoted content are highest on the first
day of the campaign, and fall considerably thereafter. However, we show that
these insights depend on the use of robust machine learning and natural language
processing techniques to gather focused, relevant datasets, and to accurately
gauge sentiment, rather than relying on the simple keyword- or frequency-based
metrics sometimes used in social media research.




%% file: intro.tex
\section{Introduction}
\label{sec:intro}

Online Social Networks (OSNs) such as Facebook, Twitter, and YouTube have emerged as highly engaging marketing and influence tools, increasingly used by advertisers to promote brand awareness and catalyze word-of-mouth marketing. Researchers have also long recognised the effectiveness of OSNs as a rich source for understanding the spread of information about the real world~\cite{youarewhattweet}. For example, Asur~\etal~\cite{DBLP:journals/corr/abs-1003-5699} analyzed Twitter messages (\emph{tweets}) to predict box-office ratings for newly released movies. Their findings shows that OSNs can be used to make quantitative predictions that outperform those of markets forecasts, by focusing on the \textit{sentiment} expressed in the tweets. 
Brands also now recognise the potential of OSNs for gathering market intelligence and insight. In 2012, Twitter announced that 79\% of people follow brands to get exclusive content.\footnote{\url{http://advertising.twitter.com/2012/05/twitter4brands-event-in-nyc.html}} This provides the opportunity for brands to participate in real-time conversations to listen to and engage users, respond to complaints and feedback, drive consumer action and broadcast content. Understanding the real engagement of the end users with the brands and their OSN presence has given rise to a number of data analytics, sentiment analysis and social media optimisation startups and academic research projects. However, the techniques required pose a number of challenges and pitfalls often ignored by researchers and analysts, and adopting a particular method naively can lead to problems. Significant progress in Natural Language Processing (NLP) and Machine Learning (ML) has produced models for topic modelling designed for social media \cite{Ritter.etal10}, and high accuracies in sentiment detection (e.g. \cite{Socher.etal13}), even with the possibility of detecting sarcasm \cite{Liebrecht.etal13}; but care still needs to be taken while using and relying on the relevant tools and techniques straight out of the box~\cite{Pollyanna@COSN13}.

In this study we present a focused case study by examining the content and volume of users' brand engagement on OSNs to determine the effect of choice of promotion channel on a brand's influence.\footnote{In this work, engagement is defined as adoption of the content by e.g., replying to a tweet, mentioning the brand name, or including the hashtag in a tweet. We are not able to measure external engagement such as sharing content on other OSNs, or clicking on the links in the tweet.} We do this by analysing the engagement level  of Twitter users, their adoption of brand hashtags, and the sentiment they express, to determine the similarities and differences between two separate advertising strategies on this network: \textit{promoted tweets}, and \textit{promoted trends}. We pose a number of questions regarding brands and advertising on OSNs: How does the sentiment for a promotion strategy spread over time? What are the engagement levels for each day of promotion? What is the engagement level (e.g. \emph{retweets} and \emph{mentions}) for promoted brands and how do these affect the sentiments expressed towards a brand?

In order to answer these questions, we use Twitter's Streaming API service to
collect engaged users' profiles and tweets in regards to promoted influences
(tweets and trends) over a busy 10 day shopping period for a selection of brands
across different industries. We observe the need to accurately filter the
resulting tweets for topical relevance, and compare simple keyword-based methods
with a discrimative Machine Learning (ML) approach. We then classify the tweets
by sentiment (positive, negative or neutral), and again compare a range of
existing methods and tools.  We then use this data to establish the driving
factors behind the success of promoted influences and differences between
advertising strategies.  For both tasks, the choice of classification method
makes a significant difference, highlighting the care that must be taken
when choosing techniques for this kind of analysis.

The rest of the paper is organized as follows: In Section~\ref{sec:related} we present the some recent related studies. In Section~\ref{sec:data} we describe our case study, dataset and its characteristics. In Section~\ref{sec:classify} we briefly discuss our sentiment analysis and text classification methodology and the challenges which only become apparent upon thorough manual inspection of the data. Section~\ref{sec:results} presents our results and the insights gained from our analysis. We conclude the paper and present potential future directions in sentiment and content analysis in Section~\ref{sec:conclusion}.

%% file: related.tex
\section{Related Work}
\label{sec:related}

\subsection*{Influence on OSNs}

Our primary interest in this work is in understanding the factors which govern
the effectiveness and influence of campaigns on OSNs. Several recent studies
have examined individuals' influence on OSNs~\cite{Cha10measuringuser}, and the
effectiveness of online advertising~\cite{Chan:2011:MLV:2047529.2047535,
  blake2013consumer}, but little attention has been paid to identifying the
driving factors behind a brand's influence on their social audience (although it
has been noted that brand names are more important online for some
categories~\cite{Degeratu200055}).  Cheung \etal~\cite{cheung2010effectiveness}
examined the way information spreads differently within social networks as
opposed to word-of-mouth (WOM) broadcasting, by focusing on electronic
word-of-mouth (eWOM), showing comprehensiveness and relevance to be the key
influences of information adoption. The closest work to ours in understanding
brands on Twitter is the study by Jansen~\etal~\cite{ASI:ASI21149}, who found
that 20\% of tweets that mentioned a brand expressed a sentiment or opinion
concerning that company, product or service. Here, we examine and compare such
mentions and sentiments across different promotion strategies available to
brands on Twitter, thus specifically investigating advertising effectiveness (see
Section~\ref{sec:data}).\footnote{Data availability limits us to effects within
  OSNs; we cannot determine effects on actual clicks or sales.}

In a study on the spread of hashtags within Twitter,
Romero~\etal~\cite{Romero:2011:DMI:1963405.1963503} used over 3 billion tweets
2009-2010 to analyze sources of variation in how the most widely used hashtags
spread within its user population. Their results suggested that the mechanism
that controls the spread of hashtags related to sports or politics tends to be
more persistent than average; repeated exposure to users who use these hashtags
affects the probability that a person will eventually use the hashtag more
positively than average. However, they only examined hashtags that succeeded in
reaching a large number of users. In regards to the focus of promoted influences
within Twitter, this raises the question; what distinguishes a promoted item
that spreads widely, possibly with positive sentiment, from one that fails to
attract attention or is associated with mainly negative sentiment? Our study
aims to answer this by examining the sentiment and spread of tweets in relation
to brands' promoted items.

\subsection*{Analysis Methods}

Sentiment analysis has been approached across many domains, including products,
movie reviews and newspaper articles as well as social media (see e.g
\cite{Pang.Lee08} for a comprehensive overview). Typically, the methods employed
depend either on existing language resources (e.g. sentiment dictionaries or
ontologies) or on machine learning from annotated datasets. The former can
provide deep insight, but are somewhat inflexible in the face of the
non-standard and rapidly changing language used on OSNs, for which few suitable
linguistic resources currently exist. The latter are more scalable and can be
trained on relevant data (e.g.~\cite{mejova2012crossing}), but generally depend
on large amounts of manual annotation (expensive and often problematic in terms
of accuracy) and in some cases the existence of grammatical resources for the
language and text domain in question (e.g.~\cite{Socher.etal13}). However, some
approaches leverage the existence of implicit labelling in the datasets
available (\emph{distant supervision}), to avoid the necessity for manual
annotation: for example, user ratings provided with movie or product
reviews~\cite{Pang:2002:TUS:1118693.1118704,Chan:2011:MLV:2047529.2047535}); or
author conventions such as emoticons and hashtags on
OSNs~\cite{Go.etal09,Pak.Paroubek10,Purver.Battersby12}). Hybrid approaches also
exist, e.g. the use of predefined sentiment dictionaries with weights learned
from data (e.g.~\cite{Thelwall:2012}).

Identifying the topic of text has also received much attention in NLP research,
with methods ranging from the use of existing topic resources or ontologies
(e.g.~\cite{malo2010semantic}) to unsupervised models for discovery of topics
(e.g.~\cite{Ritter.etal10}). The use of machine learning to detect the relevance
(or otherwise) of text to a known topic also has a long history, perhaps most
well-known in the form of Na\"{i}ve Bayes filtering for spam filtering
\cite{Sahami.etal98}.

However, research into OSN behaviour or influence sometimes ignores the spread
of sophisticated methods available. Sentiment analysis is often performed based
on defined dictionaries (e.g.~\cite{Tumasjan.etal10}), and topic identification
is often ignored, with datasets filtered purely on keywords or simple Boolean
queries.  Recently, Goncalves~\etal~\cite{Pollyanna@COSN13} examined the
difference in performance across various sentiment analysis approaches on online
text, finding significant variations. The effect of these variations in a
specific analysis problem is less clear, though: how much does the variation in
sophistication (and accuracy) of these methods actually matter?
\cite{Quercia.etal11b} compared statistical and lexicon-based methods and found
significant differences at the level of individual messages, although a
correlation at the level of their intended analysis (user profiles). Here, we
investigate the effect when considering individual advertising campaigns
(promoted items). For text relevance, we compare the use of keywords to Na\"ive
Bayes classification via Weka \cite{Hall.etal09}. For sentiment analysis, we
examine three existing and freely available tools: the widely-used Data Science
Toolkit's
\texttt{text2sentiment}\footnote{\url{http://www.datasciencetoolkit.org/}} based
on a sentiment lexicon \cite{Nielsen11}; the lexicon-based but
data-driven hybrid SentiStrength \cite{Thelwall:2012}; and a statistical
machine-learning-based approach, Chatterbox's
Sentimental\footnote{\url{http://sentimental.co/}}
(see~\cite{Purver.Battersby12}).

%% file: data.tex
\section{Data collection}
\label{sec:data}

We set up a crawler to use the Twitter Streaming API\footnote{\url{https://dev.twitter.com/docs/streaming-apis}} to collect the tweets of interest and all associated metadata (e.g., ID, username, user's social graph), with details stored in a MySQL database. In this section we briefly describe our dataset and data collection strategy. 

\subsection*{Identifying promoted brands}
Twitter distinguishes promoted tweets and trends by the use of a \emph{Promoted} tag. We collected tweets from 11 brands with an active advertising campaign during our study period, across different industry domains, ranging from entertainment to health-care. For each promoted item, the brand names was used to crawl Twitter for tweet data posted in English for a 10 day period. If the promoted item also included a hashtag, the hashtag was also included in the parameters of the crawl's GET function. This included all tweets that contained keywords such as  \texttt{@BrandName}, \texttt{\#BrandName}, \texttt{BrandName}, \texttt{\#PromotedHashtag} and other brand related terms. These parameter values were selected to keep the dataset both relevant to brand-related tweets, and also manageable for searching purposes. Followers and following information was also tracked on a daily basis for each brand.  

\begin{table}[t]
\centering
\begin{tabular}{ |l|l|l| }
\hline
\textbf{Industry}&\textbf{Promotion type}&\textbf{Brand} \\ \hline
\multirow{3}{*}{Electronics}&Promoted tweet&International CES \\
 &Promoted tweet&SONY \\
 &\textit{Promoted trend}&Nintendo UK \\
  \hline
\multirow{1}{*}{Travel}&Promoted tweet&Marriot \\
  \hline
\multirow{1}{*}{Entertainment}&Promoted tweet&BBC One \\
  \hline
\multirow{1}{*}{Automobile}&\textit{Promoted trend}&Vauxhall \\
  \hline
\multirow{1}{*}{Heath Care}&Promoted tweet&Paints like Me \\
  \hline
\multirow{3}{*}{Retail}&\textit{Promoted trend}&ASOS \\
 &\textit{Promoted trend}&PespiMax \\
 &Promoted tweet&JRebel \\
  \hline
\multirow{1}{*}{Telecomms}&\textit{Promoted trend}&O2 Network \\
  \hline
 \end{tabular}
        \vspace{-0.2cm}
\caption{Industry sectors and sample brands} 
        \vspace{-0.5cm}
\label{tab:PPer}
\end{table}

Details of the selected brands and their promoted type are provided in Table~\ref{tab:PPer}. Given that we were interested in promoted items for branding purposes, a range of different brands from different industries were selected. The aim was to include both major, and small brands when selecting promoted items. In addition, a major brand and a small brand enable a comparison of sentiment while weakly controlling for follower count.

\subsection*{Dataset}

We identified different industries' promoted items for 10 day periods between $17^{th}$ December 2012 and $7^{th}$ January 2013. We used non-parallel crawling periods in order to avoid the query limits set by the Twitter API. In total, around 180,000 individual tweets were collected by crawling Twitter continuously, excluding December $21^{st}$ 2012 when there was a 6 hour outage in the crawler API. The crawler collected tweets from around 120,000 different Twitter users engaged in spreading the promoted tweets and trends. Tweets across all topics and with no geographical limits were gathered, as long as they featured the brand's name/hashtag. When a brand contained more than one directly relevant hashtag, e.g., \texttt{\#Coke} and \texttt{\#CocaCola}, we included all the relevant hashtags.

Twitter users do often repeat their tweets to benefit from repeated exposure. However, in order to remove noise and bias in analysis caused by spam tweets, we removed users who had posted the exact same tweet more than 20 times during our measurement periods, along with their tweets. Twitter users, tweets and tweet timestamps were also cross-analysed to check for spamming accounts. In one case a single user was removed for adding over 8,000 spam tweets to the database. After manual inspection of many tweets and accounts, we are confident that nearly all spam has been removed from our dataset.

%% file: classify.tex
\section{Text Processing \& Classification}
\label{sec:classify}

In this section we present the details of our tweet classification (using ML)
and sentiment analysis (using existing NLP tools).

\subsection{Topic Classification}
\label{sec:topic}

One of the major challenges during cleaning the dataset and removing spam was
ensuring topic relevance. Our expectation was that this would not be an issue:
as in much previous work, our study is looking at all sentiment expressed
towards the brands, as long as the tweet matched the parameters of the tweet
selection as explained in Section~\ref{sec:data}. However, whilst sampling tweets for spammers, a general problem surfaced. We found that a keyword-based approach
tends to be too broad to accurately identify tweets referring to a particular brand,
\emph{O2} (a UK mobile telecommunications provider and network). Our parameters
for collecting tweets for this brand were to match tweets containing
\texttt{O2WhatWould\-YouDo} and \texttt{O2} (the hashtag being promoted was
\texttt{\#O2What\-WouldYouDo} and \texttt{@O2} is the official brand Twitter
handle). Over the 10 day period, 90,000 tweets were collected that matched
these keywords. However, examining a random sample of 200 tweets from this
dataset showed that over 70\% were not referring to the O2 Network brand; many
were referring to the \textit{``O2 Academy''} (a chain of concert venues), the
\textit{``O2 Arena''} (a dome-shaped monstrosity in London), or other senses of
\textit{`O2'} such as oxygen. We also noticed that Twitter users have recently
established a new way of using the letter sequence \textit{`O2'} as a
replacement for the letters \textit{`to'}: e.g. ``\texttt{@CokeWave\_Thang What
  Picture You Want Me O2 Put As My BackGround}'', ``\texttt{what im\\goin o2 do
  o2day}''. Experiments with boolean combinations of \texttt{O2} with other
keywords were not successful. A major challenge therefore becomes to filter out
non-brand-related tweets automatically: the problem is not trivial, given the
variability and unpredictability of language, vocabulary and spelling on
Twitter, and the short length of tweets (up to 140 characters); and manual
removal of approximately 70\% of large datasets is prohibitively
labour-intensive.


We therefore approached this as a text classification problem and investigated
various supervised machine learning approaches using the Weka toolkit
\cite{Hall.etal09}. First, we performed a pilot study over a 200-tweet
development set to determine a suitable feature representation and
classification method; the data was manually labelled as O2-related or otherwise
to give a binary decision problem. We tested a variety of classifiers including
Naive Bayes, Naive Bayes Multinomial, ID3, IBK and J48 decision trees; features
were based on the tweet text using a standard bag-of-words representation (see
e.g. \cite{Manning.Schutze99}) with various scaling methods,\footnote{We used
  Weka's \texttt{StringToWordVector} filter for text feature extraction and
  scaling.} with the addition of user ID and date of tweet. Given the small size
of the dataset, we restricted the feature space to be based on the most common
100 words. We also tested using a simple manual keyword-based filter to remove some
common negative instances (using keywords \textit{arena, academy}, etc) before
training (see ``manually filtered'' results in the figures). Tests were
performed using ten-fold cross-validation in order to simulate performance on
unseen data. Best performance (overall accuracy) was obtained using only
bag-of-words text features, with stopwords removed and a TF-IDF weighting, after
manual filtering. The best performing classifiers in cross-validation were J48
and Naive Bayes (NB), with 71\% and 91\% accuracy respectively. We then compared
their performance on a held-out test set: the NB model outperformed the J48
model with 84\% accuracy compared to 71\% for J48, with training and prediction
also noticeably faster for NB (the tree structure of the J48 model made it very
slow with larger training sets).

\begin{figure}[ht!]
  \centering
  \includegraphics[width=0.5\textwidth]{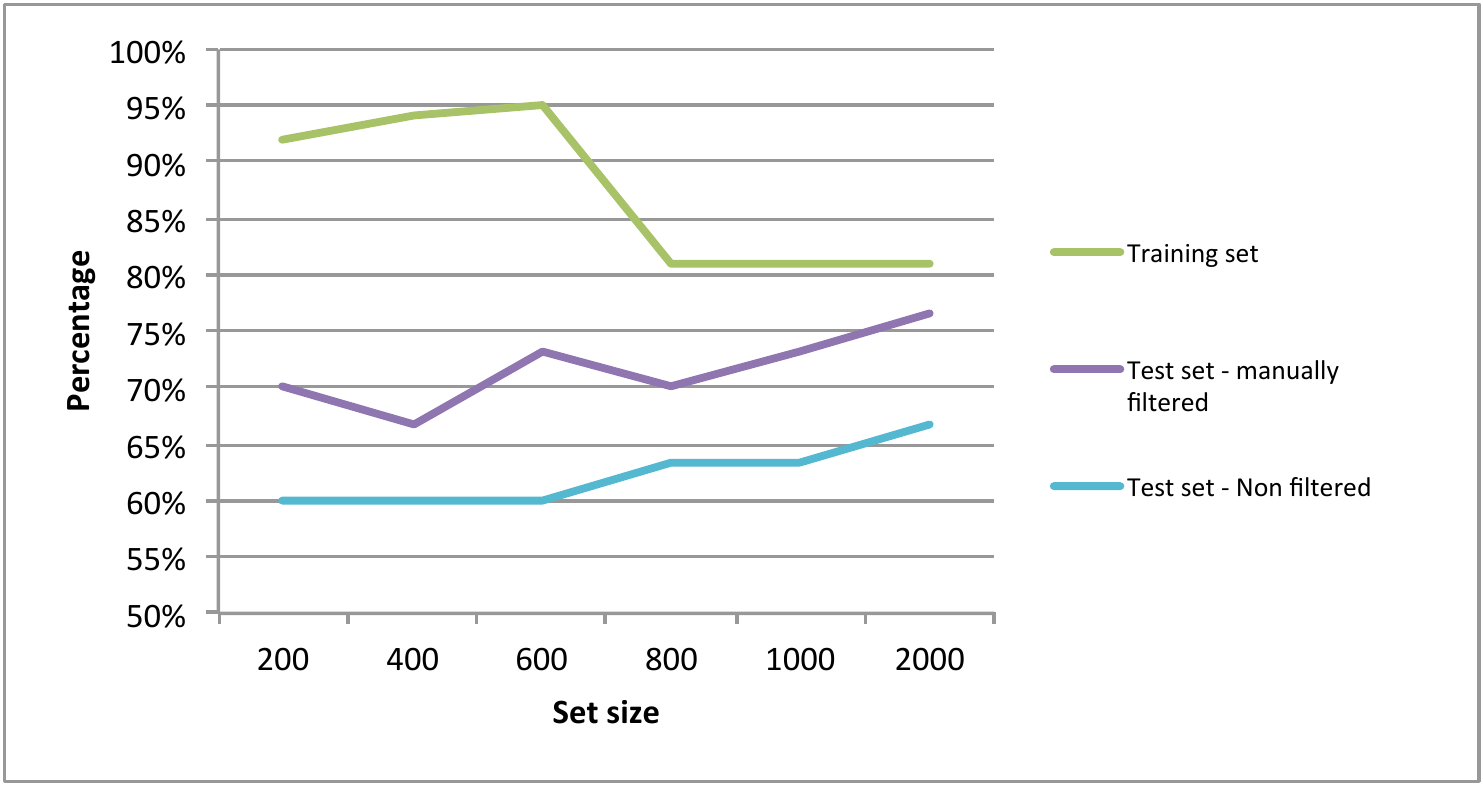}
  \vspace{-0.5cm}
  \caption{NB accuracy with increasing training data.}\label{fig:learning-curve}
  \vspace{-0.25cm}
\end{figure}

\begin{figure}[t!!]
\centering
  \includegraphics[width=3.3in,height=2.5in]{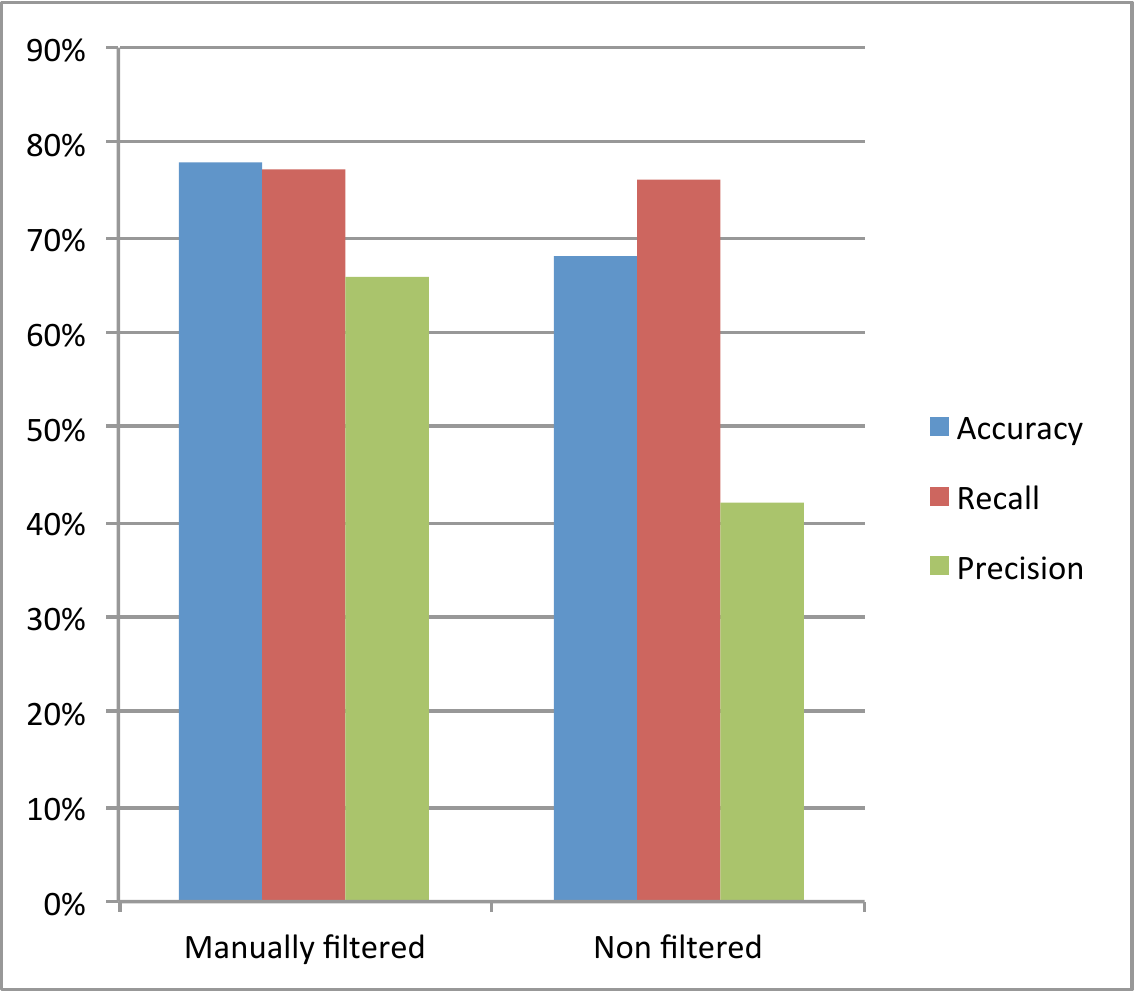}
 \caption{Classification results using Naive Bayes.\label{fig:NBres}}
\end{figure}

\begin{figure}[t!!]
\centering
  \includegraphics[width=3.3in, height=2.5in]{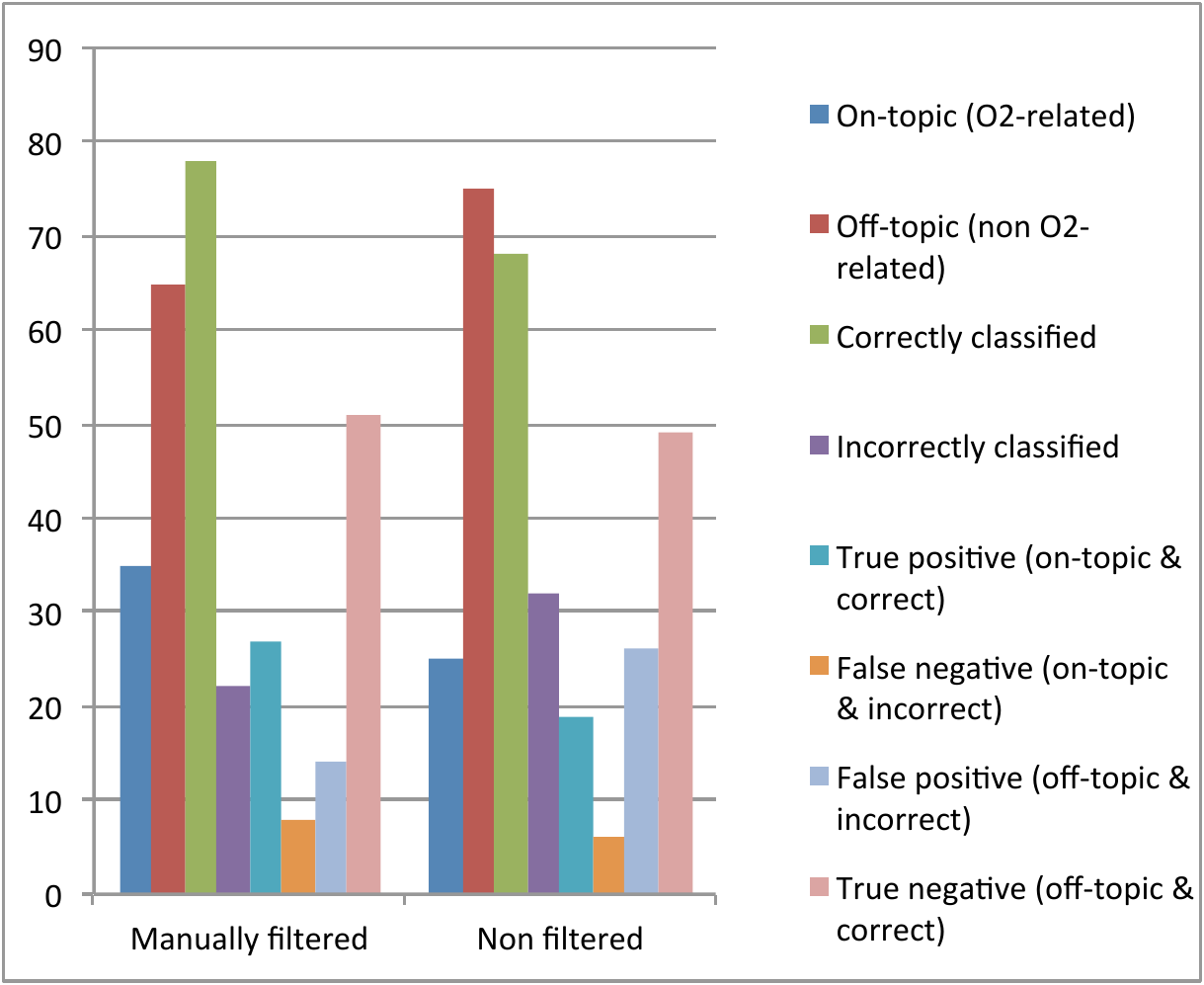}
 \caption{Classification details per class using Naive Bayes.\label{fig:NBres2}}
\end{figure}

To determine a suitable training set size, we then varied the training set while
testing performance on a held-out test dataset of 30 manually labelled
tweets. Increasing training set size improved performance (see
Figure~\ref{fig:learning-curve}): we tested up to a 2,000-tweet training set;
while the curve suggests performance may improve beyond this point, the accuracy
on the held-out test set is approaching that on the training set so large
improvements are unlikely. The NB classifier trained on 2,000 tweets was
therefore used for the experiments below. Figure~\ref{fig:NBres} shows results
when tested on a larger, unseen, randomly selected test set of 100 tweets; the
version with manual filtering achieves 78\% accuracy, 77\% recall and 66\%
precision. Figure~\ref{fig:NBres2} gives details of the per-class predictions:
without manual filtering, false positives are more common than false negatives
(i.e. too much irrelevant data is slipping through); levels are much closer with
filtering.

%

\subsection{Sentiment Analysis}
\label{sec:sentiment}

Having identified tweets with relevant content, we now required a method for
sentiment analysis -- determining the positive or negative stance of the writer.
As discussed in Section~\ref{sec:related} above, many methods for sentiment
detection exist, with the major distinction being between lexicon-based and
machine learning-based approaches. We examined existing tools for Twitter
sentiment analysis using both of these approaches in order to determine the most
suitable for our data.

As a baseline lexicon-based tool we used the freely available Data Science
Toolkit\footnote{\url{http://www.datasciencetoolkit.org/}}. The sentiment
analyser is based on a sentiment lexicon \cite{Nielsen11}; we therefore
anticipate its coverage to be low but take it to be representative of
commonly-used lexicon-based approaches.

For a more robust tool for comparison, we examined two alternatives.  As a
hybrid lexicon/machine-learning tool we chose
SentiStrength~\cite{Thelwall:2012}. This method uses a predetermined list of
words commonly associated with negative or positive sentiment, which are given
an empirically determined weight (learned from data); new texts are classified
by summing the weights of the words they
contain. Thelwall~\etal~\cite{Thelwall:2012} report accuracy on Twitter data of
63.7\% for positive sentiment and 67.8\% for negative when predicting ratings on
a 1-5 scale, and accuracies near 95\% when predicting a simple binary
positive/negative label. However, even though their word lists and weightings
are determined for OSN data (including Twitter), this approach may suffer when
faced with social text with new words, unexpected spellings and
context-dependent language and meaning (see~\cite{6215730}).

For a purely ML-based option we used Chatterbox's Sentimental
API,\footnote{\url{http://mashape.com/sentimental/sentiment-analysis-for-social-media}}
based on statistical machine learning over large, distantly labelled
datasets~\cite{Purver.Battersby12}. This data-based approach means it might be
expected to handle slang, errorful or abbreviated text better. Purver \&
Battersby \cite{Purver.Battersby12} report accuracies approaching 80\% using a
similar technique on smaller datasets; Chatterbox report 83.4\% accuracy in an
independent study.\footnote{See
  \url{http://content.chatterbox.co/Sentiment\%20Analysis\%20Case\%20Study\%20-\%20Chatterbox\%20and\%20IDL.pdf}.}

Before applying the sentiment analysis tool, and in order to compare the two
approaches, a few hundred random tweets were selected from the database and were
read and manually labelled for positive or negative sentiment, and both tools
were tested on the resulting set. Results showed accuracy below 50\% for the
lexicon-based Data Science Toolkit, 63\% for the hybrid SentiStrength approach,
and 84\% for the ML-based Chatterbox approach. Error analysis showed one
significant source of the latter difference to be sentiment expressed in
hashtags (e.g. the negative \texttt{\#shambles}), which were detected better by
the ML-based approach, presumaby due to their absence from SentiStrength's
predetermined lexicon. We therefore use Chatterbox as the ``robust'' tool in our
experiments below, and compare to the Data Science Toolkit as a purely
lexicon-based baseline.

%% file: results.tex
\section{Results}
\label{sec:results}


\begin{figure}[t!!]
\centering
  \includegraphics[width=3.3in, height=2.5in]{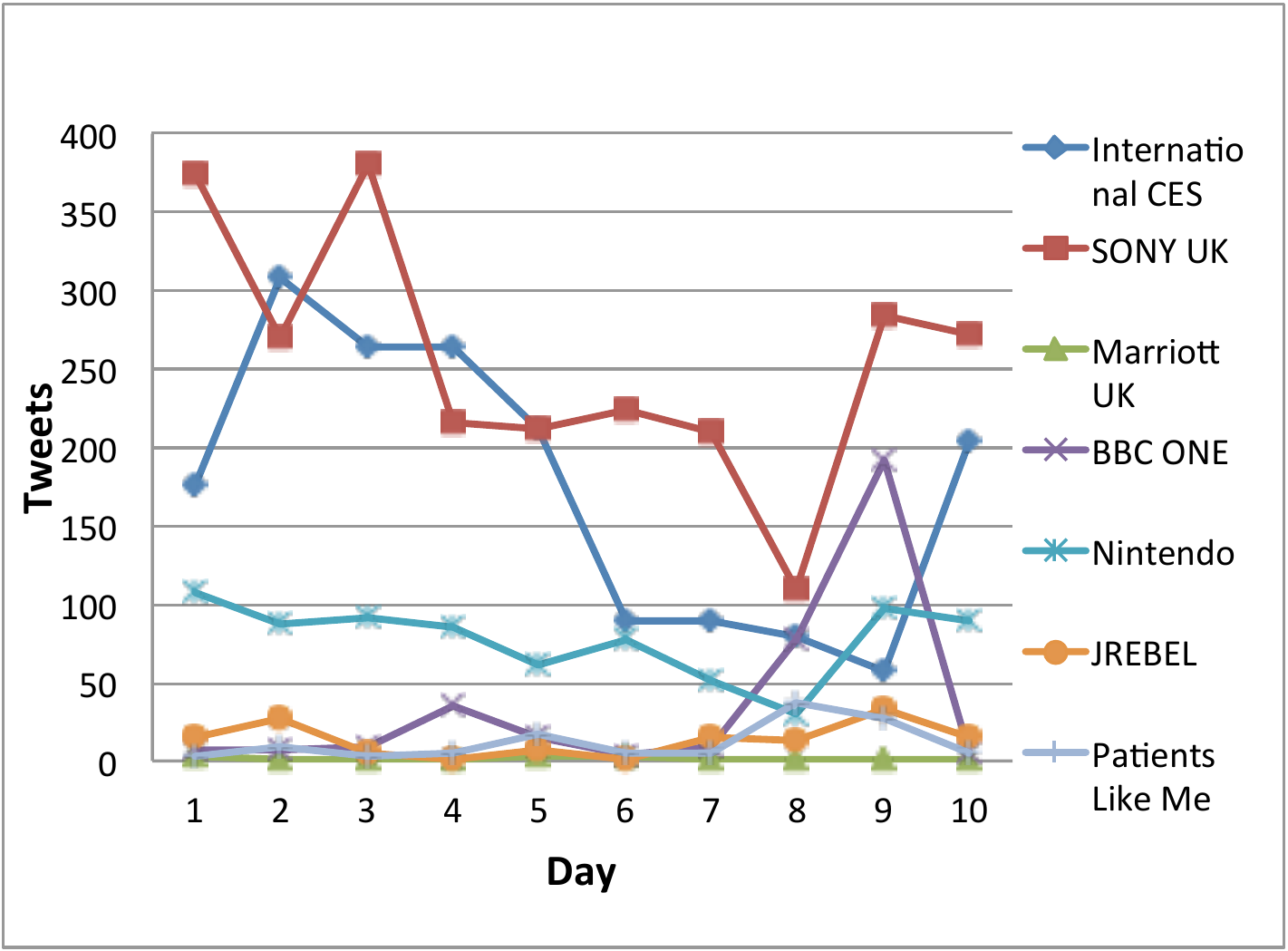}
 \caption{Distribution of promoted tweets volumes over time.\label{fig:disttweets}}
\end{figure}

\begin{figure}[t!!]
\centering
  \includegraphics[width=3.3in,height=2.5in]{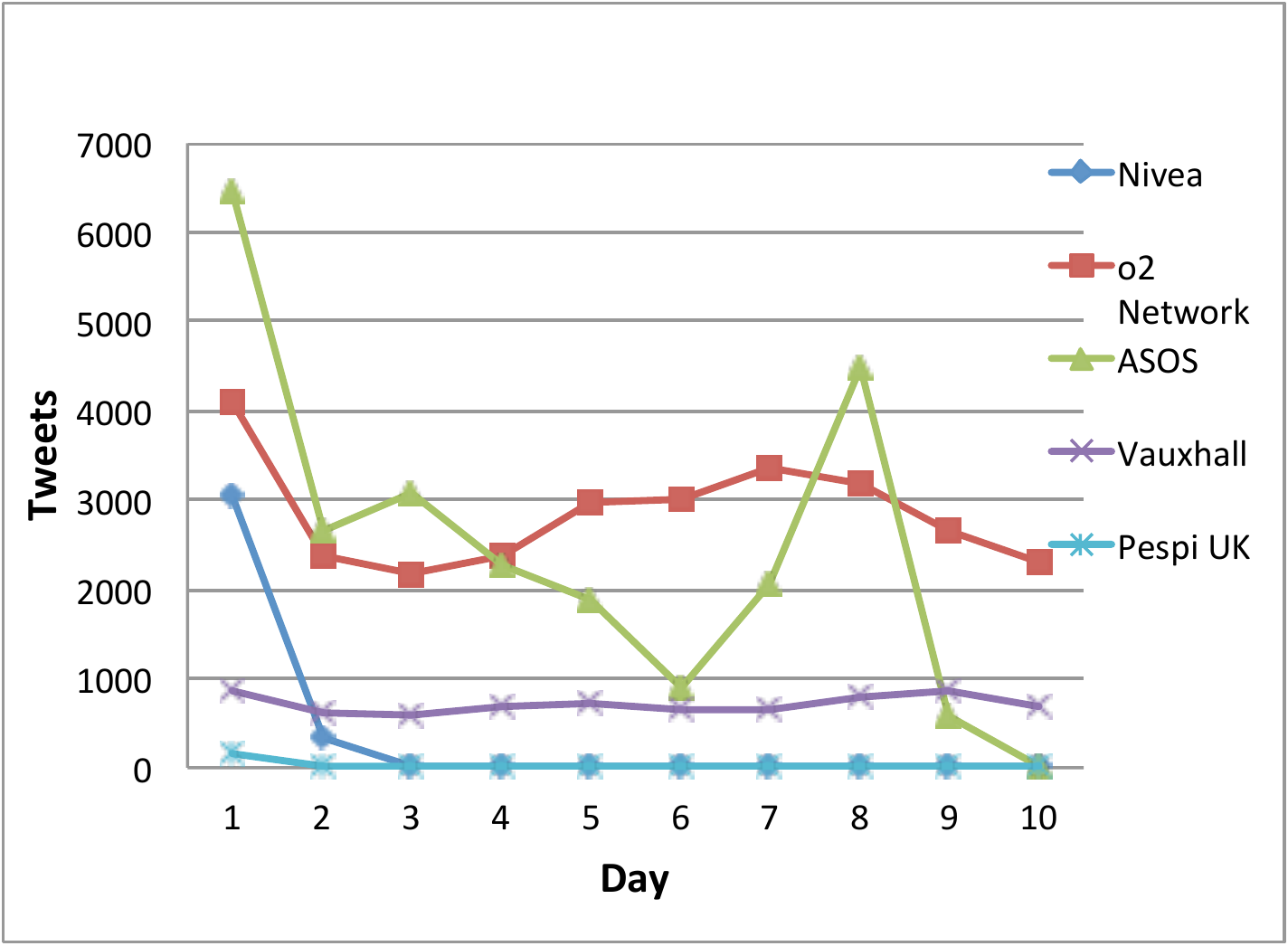}
 \caption{Distribution of promoted trends volumes over time.\label{fig:dists}}
\end{figure}

\subsection*{Response volume over time}



To examine the spread of engagement for each promoted item over the 10 day
period, we analysed the volume of unique tweets each day in response to each
promoted item, then averaged the results across all
brands. Figures~\ref{fig:disttweets} and~\ref{fig:dists} display the distribution of this volume in
response to \emph{promoted tweets}~(\ref{fig:disttweets}) and \emph{promoted
  trends}~(\ref{fig:dists}) per brand.
On average, promoted trends led to much higher response volumes. However, the
highest percentages of \emph{mentions} within responses were from promoted
tweets, where an average of 18\% of tweets each day included an `@' mention to
the brand; promoted trends had an average of only 15\% mentions per day. This indicates that for a brand to successfully
engage users in the content of the promoted item, a promoted tweet is better for
this purpose. For example, out of the O2 Network's $\sim$30,000 tweets, 7,965 included an `@'mention to the brand (25\%).  

Results confirmed that the greatest percentage of engagement for a brand's
promoted item takes place on the first day of promotion.
On average, 24\% of engagements around the
promoted item take place on the first day. The effect is most
pronounced for \emph{promoted trends}, with 34\% of engagement on average on the
first day of promotion, after which the engagement falls dramatically by an
average of 25\% to 9\% by day two and continues to fall thereafter, even if the item is promoted for several days.  For
\emph{promoted tweets}, the effect is less pronounced: 19\% of the engagement
takes place on the first day of promotion, with engagement decreasing by 8\% by
the second day of promotion. However, it does not continue on a steady decline
thereafter, but it rises and falls over the next 8 days, although never again
reaching the peak of the first day of promotion. This could be due to the fact
that a promoted tweet is usually promoted for several days on Twitter where it
occasionally appears at the top of different user's timeline were users are
repeatedly exposed to the item. This finding can be said to conform to
Romero~\etal's theory of \textit{repeated
  exposure}~\cite{Romero:2011:DMI:1963405.1963503}.\footnote{Also see
  \small{http://advertising.twitter.com/2013/03/Nielsen-Brand-Effect-for-Twitter-How-Promoted-Tweets-impact-brand-metrics.html}} They found that repeated exposure to a hashtag within Twitter had a significant marginal effect on the probability of adoption of that hashtag.

In general, though, these results show that adoption of a promoted item is not
a slow gradual shift over several days (as might be assumed) but rather an
immediate incline when exposure to the item is new to users.

\subsection*{Effects on user sentiment}

\begin{figure}[h!]
  \includegraphics[width=\columnwidth]{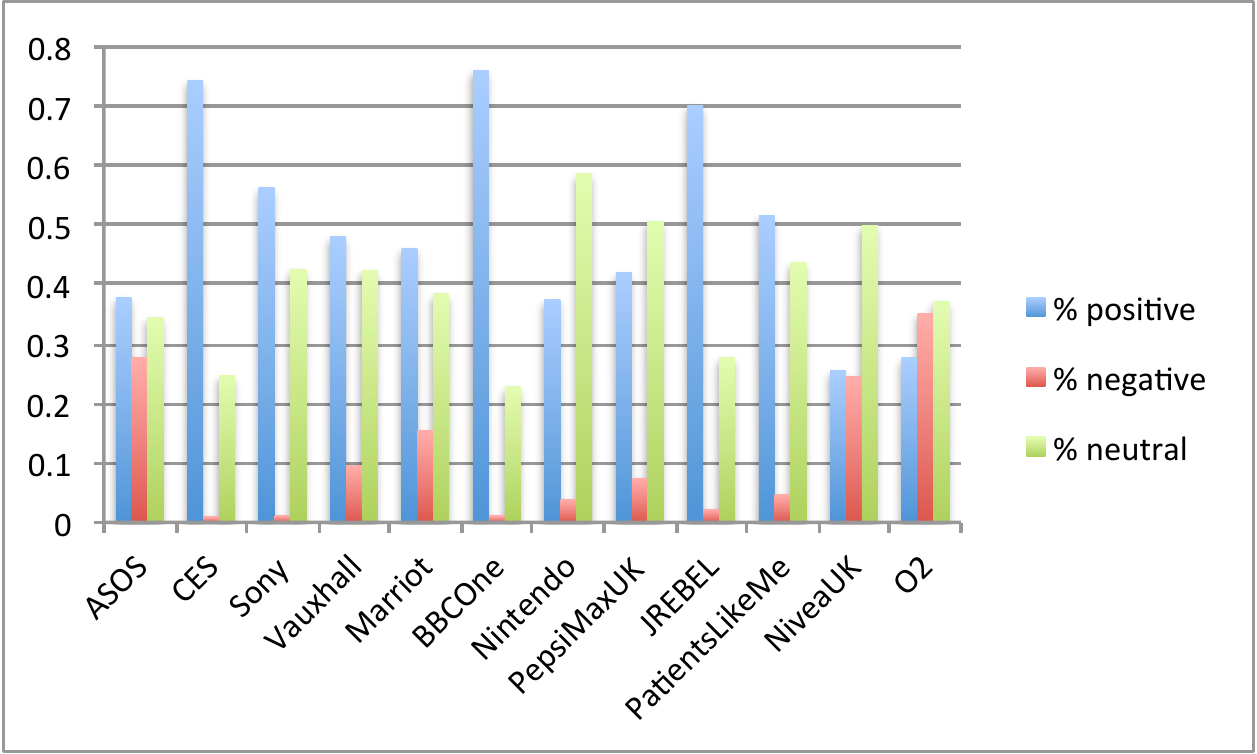}
  \vspace{-0.5cm}
  \caption{Sentiment analysis by brand - machine learning}\label{fig:brandsentimentml}
\end{figure}

\begin{figure}[h!]
  \includegraphics[width=\columnwidth]{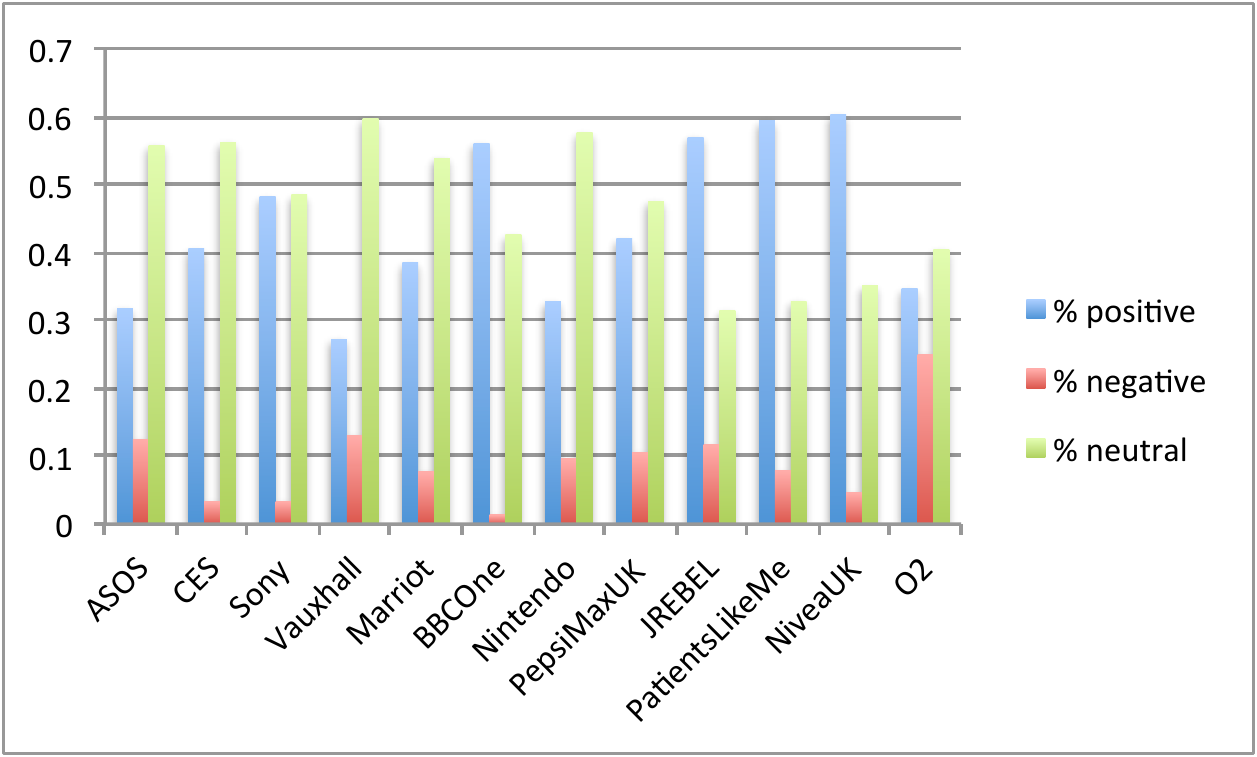}
  \vspace{-0.5cm}
  \caption{Sentiment analysis by brand - keywords}\label{fig:brandsentimentkey}
\end{figure}

The sentiment breakdown for each promoted brand item can be observed in
Figures~\ref{fig:brandsentimentml} and ~\ref{fig:brandsentimentkey}, with
Figure~\ref{fig:brandsentimentml} showing the results obtained using our chosen
machine learning method and Figure~\ref{fig:brandsentimentkey} those obtained
using a keyword-based method (see section~\ref{sec:classify} above). We observe
that in most cases, the percentage of positive sentiment was higher than that of
negative and neutral for promoted items. Notable exceptions are the results for
two brands, NiveaUK and O2, where neutral and/or negative levels outweigh
positive; the ASOS brand also shows little difference between negative and
positive levels. However, comparison of the figures that would have been gained
using a keyword-based approach (Figure~\ref{fig:brandsentimentkey}) shows
misleading results in precisely these interesting cases: apparent positive
levels are higher than negative in all cases. Neutral cases also appear much
more common; this is due to the low coverage of the keyword lexicon causing
large numbers of results with apparently zero sentiment. Use of the more
accurate tool (as objectively assessed -- see section~\ref{sec:classify})
therefore does appear crucial.

On average, across all brands (promoted tweets and trends), the average
percentage of tweets and retweets\footnote{We assume that retweeting users share
  the same sentiment as the original tweet.}  which contained a positive
sentiment is 50\%, that which contained a negative sentiment is 12\%, and 38\%
of tweets had a neutral tone.

\begin{figure}[t!]
    \centering
    \includegraphics[width=3.3in, height=2.5in]{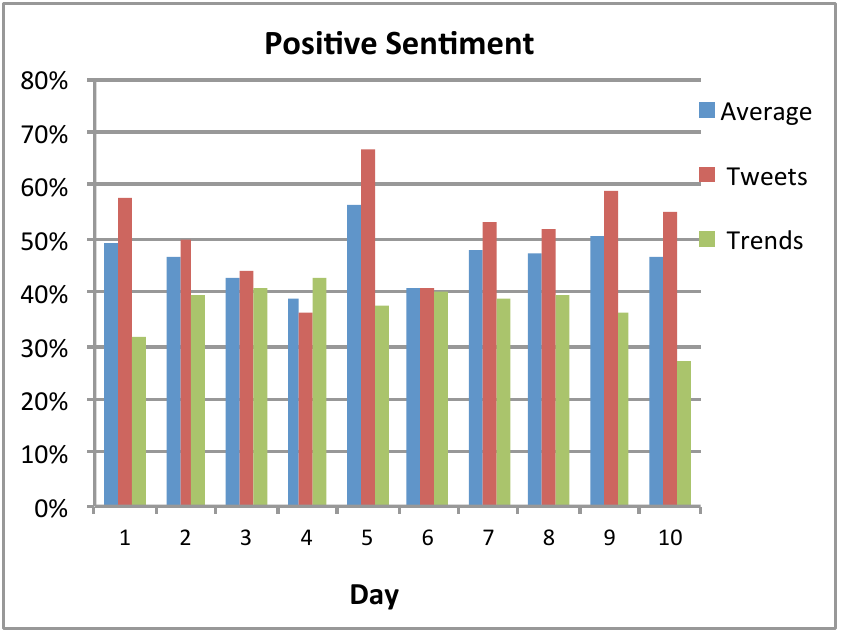}
    \caption{Positive sentiment distribution over time}
     \label{fig:positivesentiment}
\end{figure}

\begin{figure}[t!]    
	\centering
    \includegraphics[width=3.3in, height=2.5in]{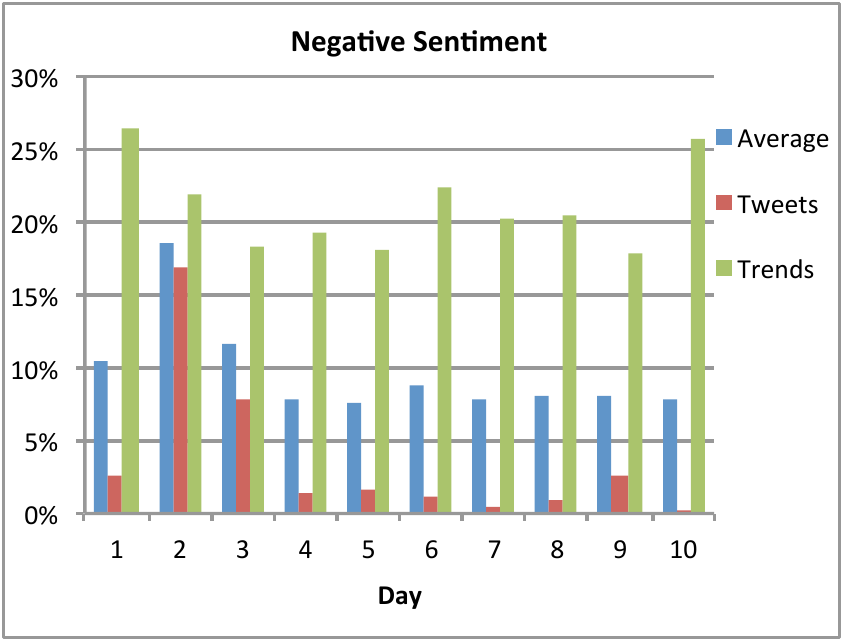}
    \caption{Negative sentiment  distribution over time.}
 \label{fig:negativesentiment}
\end{figure}

Figures~\ref{fig:positivesentiment} and~\ref{fig:negativesentiment} then show
the distribution of positive and negative sentiments in this response traffic
over time. On average, positive sentiment outweighs negative sentiment; on the
first day, 49\% of the tweets were positive. In general, \emph{promoted tweets}
lead to more positive sentiment and less negative sentiment than \emph{promoted
  trends}.

In total, 47\% of tweets relating to a \emph{promoted tweet} are positive in
sentiment. Day one received the highest percentage of positive sentiment tweets
(58\%); positive sentiment then continues to dominate over the 10 day period,
never falling below 36\% of the tweets.  Examining \emph{promoted trends}, we
found that, on average, only 37\% of tweets relating to a promoted trend
contained a positive sentiment. On the first day of promotion, 26\% of tweets
expressed a negative sentiment, 32\% expressed a positive sentiment and 42\%
expressed no sentiment at all. This shows that Twitter users do not tweet as
positively about a promoted trend as they would about a promoted tweet. Instead,
a large proportion of tweets relating to a promoted trend contained no emotional
words, or if they did, the positive and negative sentiments balanced each other
out. They generally contained just the promoted hashtag or generally had an
objective, matter-of-fact tone (e.g., - ``\texttt{Get 3G where I
live... \#O2WhatWouldYouDo}'').

Taken together with the analysis of engagement volume, these results show that
when an item is promoted, the brand and the item get adopted immediately and
regarded quite positively by the engaged users. Twitter users welcome the
promoted item on Twitter, which has a positive effect on the tweets
expressed. The engagement level reduces to an average of 10\% of the total
tweets on day two, when the item is no longer being promoted, or is no longer
seen as ``new and interesting''. However, on average, the positive sentiment
expressed still outperforms that of negative sentiment and neutral sentiment
each day.

\subsection*{Effect of hashtags on engagement and sentiment}


\begin{figure}[t!!]
\centering
  \includegraphics[width=3.3in, height=2.5in]{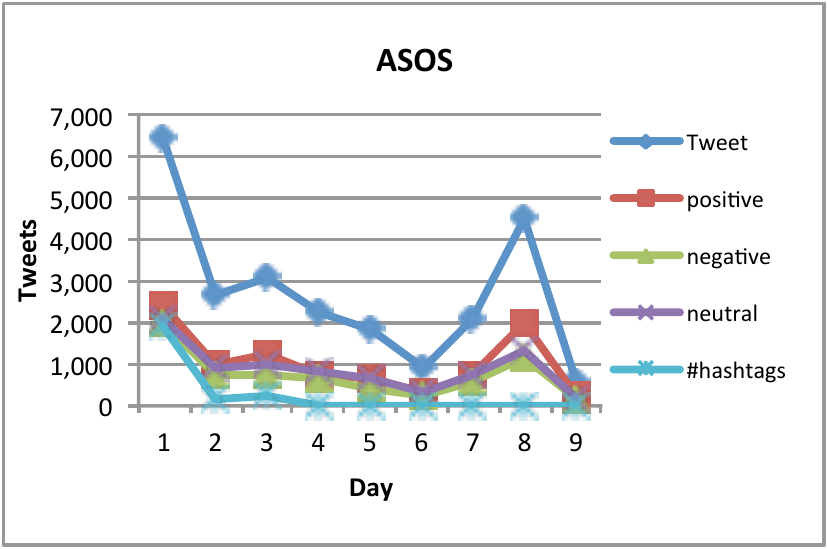}
 \caption{Hashtag related engagements for ASOS.\label{fig:hashtag-asos}}
\end{figure} 

\begin{figure}[t!!]
\centering
  \includegraphics[width=3.3in, height=2.5in]{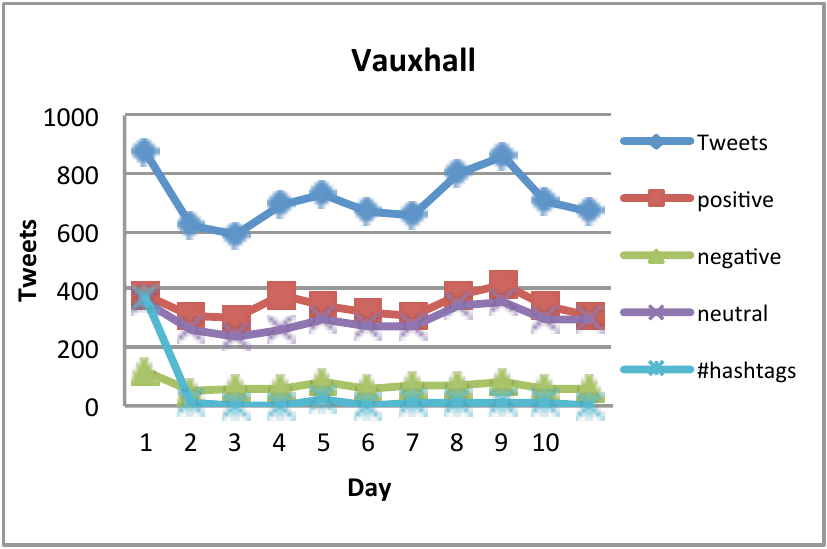}
 \caption{Hashtag related engagements for Vauxhall.\label{fig:hashtag-vaux}}
\end{figure} 

We then performed two example case studies, using the ASOS and Vauxhall brands,
to examine the use of hashtags within promoted items. Figures~\ref{fig:hashtag-asos} and~\ref{fig:hashtag-vaux} show the results. ASOS promoted a trend, \#AsosSale, on the $19^{th}$ and $20^{th}$ of
December to highlight their Boxing Day sale on the $26^{th}$ of December (day 8
of data collection).  Although the \textit{promoted} hashtag was virtually discarded by
day two of data collection, we found that user engagement (use of hashtag, mentions and tweets) for the forthcoming sale
continued. This trend is also apparent in Vauxhall's tweet volumes for their
sale which stated on the $27^{th}$ of December (day one of promotion), and ended
the day after our 10 day data collection period. The engagement for Vauxhall
remained at a consistent level throughout the event (see Figures~\ref{fig:dists} and \ref{fig:hashtag-vaux}), despite the rapid drop-off in use of the promoted hashtag.


%% file: conclusion.tex
\section{Conclusions \& Future directions}
\label{sec:conclusion}

In this paper we present a measurement-driven study of the effects of promoted
tweets and trends on Twitter on the engagement level of users, using a number of
ML and NLP techniques in order to detect relevant tweets and their
sentiments. Our results indicate that use of accurate methods for sentiment
analysis, and robust filtering for topical content, is crucial. Given this, we
then see that promoted tweets and trends differ considerably in the form of
engagement they produce and the overall sentiment associated with them. We found
that promoted trends lead to higher engagement volumes than promoted tweets.
However, although promoted tweets obtain less engagement than promoted trends,
their engagement forms are often more brand inclusive (more direct mentions);
and while engagement volumes
drop for both forms of promoted items after the first day, this effect is
less pronounced for promoted tweets.
We also found that although the volume of tweets is highest in promoted trends,
they do not lead to the same level of positive sentiment that promoted tweets
do. Hence advertisers should carefully assess the trade-offs between high level
of engagement, drop-off rate, direct mentions, and positive user sentiment.

In the next stage of this study we will investigate the effect of individuals'
influence on the take-up of promoted tweets and trends by their social graph. We
will investigate new data at finer granularity (hourly) for events that are
time-sensitive, such as major concert ticket sales. This is our first attempt at understanding this space. The advertising campaigns have very different structure and we need to understand these in details. Promoted trends typically stay on the trends list for a day, and promoted tweets are selectively shown to a subset of users for a period of time selected by the advertiser. Without accounting for such nuances, broad statements on the impact of the two forms of advertising are not conclusive. However in this paper we focussed on insights in using sentiment analysis methods and accurate data labelling. We believe our findings could provide a new insight for social network marketing and advertisements
strategies, in addition to comparing different methods of classifying and filtering relevant content.


%% file: promotweet.bbl
\begin{thebibliography}{10}

\bibitem{DBLP:journals/corr/abs-1003-5699}
S.~Asur and B.~A. Huberman.
\newblock Predicting the future with social media.
\newblock {\em CoRR}, abs/1003.5699, 2010.

\bibitem{blake2013consumer}
T.~Blake, C.~Nosko, and S.~Tadelis.
\newblock Consumer heterogeneity and paid search effectiveness: A large scale
  field experiment.
\newblock 2013.

\bibitem{Cha10measuringuser}
M.~Cha, H.~Haddadi, F.~Benevenuto, and K.~P. Gummadi.
\newblock Measuring user influence in twitter: The million follower fallacy.
\newblock In {\em in ICWSM Õ10: Proceedings of international AAAI Conference on
  Weblogs and Social}, 2010.

\bibitem{Chan:2011:MLV:2047529.2047535}
T.~Y. Chan, C.~Wu, and Y.~Xie.
\newblock Measuring the lifetime value of customers acquired from google search
  advertising.
\newblock {\em Marketing Science}, 30(5):837--850, Sept. 2011.

\bibitem{cheung2010effectiveness}
C.~M. Cheung and D.~R. Thadani.
\newblock The effectiveness of electronic word-of-mouth communication: A
  literature analysis.
\newblock {\em Proceedings of the 23rd Bled eConference eTrust: Implications
  for the Individual, Enterprises and Society}, 2010.

\bibitem{Degeratu200055}
A.~M. Degeratu, A.~Rangaswamy, and J.~Wu.
\newblock Consumer choice behavior in online and traditional supermarkets: The
  effects of brand name, price, and other search attributes.
\newblock {\em International Journal of Research in Marketing}, 17(1):55 -- 78,
  2000.

\bibitem{Go.etal09}
A.~Go, R.~Bhayani, and L.~Huang.
\newblock {T}witter sentiment classification using distant supervision.
\newblock Master's thesis, Stanford University, 2009.

\bibitem{Pollyanna@COSN13}
P.~Goncalves, M.~Araujo, F.~Benevenuto, and M.~Cha.
\newblock Comparing and combining sentiment analysis methods.
\newblock In {\em Proceedings of the 1st ACM Conference on Online Social
  Networks (COSN'13)}, 2013.

\bibitem{Hall.etal09}
M.~Hall, E.~Frank, G.~Holmes, B.~Pfahringer, P.~Reutemann, and I.~H. Witten.
\newblock The {WEKA} data mining software: An update.
\newblock {\em {SIGDKDD} Explorations}, 11(1):10--18, 2009.

\bibitem{ASI:ASI21149}
B.~J. Jansen, M.~Zhang, K.~Sobel, and A.~Chowdury.
\newblock Twitter power: Tweets as electronic word of mouth.
\newblock {\em Journal of the American Society for Information Science and
  Technology}, 60(11):2169--2188, 2009.

\bibitem{Liebrecht.etal13}
C.~Liebrecht, F.~Kunneman, and A.~Van~den Bosch.
\newblock The perfect solution for detecting sarcasm in tweets \#not.
\newblock In {\em Proceedings of the 4th Workshop on Computational Approaches
  to Subjectivity, Sentiment and Social Media Analysis}, pages 29--37, Atlanta,
  Georgia, June 2013. Association for Computational Linguistics.

\bibitem{malo2010semantic}
P.~Malo, P.~Siitari, O.~Ahlgren, J.~Wallenius, and P.~Korhonen.
\newblock Semantic content filtering with {W}ikipedia and ontologies.
\newblock In {\em {IEEE} International Conference on Data Mining Workshops
  ({ICDMW})}, pages 518--526. IEEE, 2010.

\bibitem{Manning.Schutze99}
C.~Manning and H.~Sch\"utze.
\newblock {\em Foundations of Statistical Natural Language Processing}.
\newblock MIT Press, 1999.

\bibitem{mejova2012crossing}
Y.~Mejova and P.~Srinivasan.
\newblock Crossing media streams with sentiment: Domain adaptation in blogs,
  reviews and twitter.
\newblock {\em Proc. ICWSM}, 2012.

\bibitem{6215730}
A.~Naradhipa and A.~Purwarianti.
\newblock Sentiment classification for indonesian message in social media.
\newblock In {\em Cloud Computing and Social Networking (ICCCSN), 2012
  International Conference on}, pages 1--5, 2012.

\bibitem{Nielsen11}
F.~{\AA}. Nielsen.
\newblock A new {ANEW}: evaluation of a word list for sentiment analysis in
  microblogs.
\newblock In {\em Proceedings of the {ESWC2011} Workshop `Making Sense of
  Microposts': Big things come in small packages}, pages 93--98, May 2011.

\bibitem{Pak.Paroubek10}
A.~Pak and P.~Paroubek.
\newblock {T}witter as a corpus for sentiment analysis and opinion mining.
\newblock In {\em Proceedings of the 7th conference on International Language
  Resources and Evaluation}, 2010.

\bibitem{Pang.Lee08}
B.~Pang and L.~Lee.
\newblock Opinion mining and sentiment analysis.
\newblock {\em Foundations and Trends in Information Retrieval},
  2(1--2):1--135, 2008.

\bibitem{Pang:2002:TUS:1118693.1118704}
B.~Pang, L.~Lee, and S.~Vaithyanathan.
\newblock Thumbs up?: sentiment classification using machine learning
  techniques.
\newblock In {\em Proceedings of the ACL-02 conference on Empirical methods in
  natural language processing - Volume 10}, EMNLP '02, pages 79--86,
  Stroudsburg, PA, USA, 2002. Association for Computational Linguistics.

\bibitem{youarewhattweet}
M.~J. Paul and M.~Dredze.
\newblock You are what you tweet: Analyzing twitter for public health.
\newblock In {\em Int'l Conference on Weblogs and Social Media, ICWSM}, 2011.

\bibitem{Purver.Battersby12}
M.~Purver and S.~Battersby.
\newblock Experimenting with distant supervision for emotion classification.
\newblock In {\em Proceedings of the 13th Conference of the European Chapter of
  the Association for Computational Linguistics ({EACL})}, pages 482--491,
  Avignon, France, Apr. 2012. Association for Computational Linguistics.

\bibitem{Quercia.etal11b}
D.~Quercia, J.~Ellis, L.~Capra, and J.~Crowcroft.
\newblock Tracking ``gross community happiness'' from tweets.
\newblock Technical Report RN/11/20, University College London, Nov. 2011.

\bibitem{Ritter.etal10}
A.~Ritter, C.~Cherry, and B.~Dolan.
\newblock Unsupervised modeling of {T}witter conversations.
\newblock In {\em Human Language Technologies: The 2010 Annual Conference of
  the North American Chapter of the Association for Computational Linguistics},
  pages 172--180, Los Angeles, California, June 2010. Association for
  Computational Linguistics.

\bibitem{Romero:2011:DMI:1963405.1963503}
D.~M. Romero, B.~Meeder, and J.~Kleinberg.
\newblock Differences in the mechanics of information diffusion across topics:
  idioms, political hashtags, and complex contagion on twitter.
\newblock In {\em Proceedings of the 20th international conference on World
  wide web}, WWW '11, pages 695--704, New York, NY, USA, 2011. ACM.

\bibitem{Sahami.etal98}
M.~Sahami, S.~Dumais, D.~Heckerman, and E.~Horvitz.
\newblock A {B}ayesian approach to filtering junk email.
\newblock In {\em {AAAI} Workshop on Learning for Text Categorization},
  Madison, WI, July 1998.
\newblock AAAI Technical Report WS-98-05.

\bibitem{Socher.etal13}
R.~Socher, A.~Perelygin, J.~Y. Wu, J.~Chuang, C.~D. Manning, A.~Y. Ng, and
  C.~Potts.
\newblock Recursive deep models for semantic compositionality over a sentiment
  treebank.
\newblock In {\em Proceedings of the Conference on Empirical Methods in Natural
  Language Processing ({EMNLP})}, 2013.
\newblock To appear.

\bibitem{Thelwall:2012}
M.~Thelwall, K.~Buckley, and G.~Paltoglou.
\newblock Sentiment strength detection for the social web.
\newblock {\em J. Am. Soc. Inf. Sci. Technol.}, 63(1):163--173, Dec. 2012.

\bibitem{Tumasjan.etal10}
A.~Tumasjan, T.~O. Sprenger, P.~G. Sandner, and I.~M. Welpe.
\newblock Predicting elections with twitter: What 140 characters reveal about
  political sentiment.
\newblock In {\em Proceedings of the Fourth International AAAI Conference on
  Weblogs and Social Media}, 2010.

\end{thebibliography}
